\documentclass[twocolumn,showpacs,aps,prb,superscriptaddress,floatfix]{revtex4}
\usepackage{graphicx}
\usepackage{dcolumn}
\usepackage{bm}

\usepackage{color}

\begin{document}

\preprint{}

\title{Lightly stuffed pyrochlore structure of single-crystalline Yb2Ti2O7 grown by the optical floating zone technique}

\author{K.A. Ross}
\affiliation{Department of Physics and Astronomy, McMaster University,
Hamilton, Ontario, L8S 4M1, Canada}
\author{Th. Proffen} 
\affiliation{Los Alamos National Laboratory, Los Alamos, New Mexico 87545, USA}
\affiliation{Spallation Neutron Source, Experimental Facilities Division, Oak Ridge National Laboratory, P.O. Box 2008, Oak Ridge, TN 37831, USA}
\author{H.A. Dabkowska} 
\affiliation{Brockhouse Institute for Materials Research, McMaster University, Hamilton, Ontario, L8S 4M1, Canada}

\author{J. A. Quilliam}
\affiliation{Department of Physics and Astronomy and Guelph-Waterloo Physics Institute, University of Waterloo, Waterloo, ON, N2L 3G1, Canada}
\affiliation{Institute for Quantum Computing, University of Waterloo, Waterloo, ON, N2L 3G1, Canada}
\affiliation{Laboratoire de Physique des Solides, UniversitŽ Paris-Sud 11, UMR CNRS 8502, F-91405 Orsay, France}

\author{L.R. Yaraskavitch}
\affiliation{Department of Physics and Astronomy and Guelph-Waterloo Physics Institute, University of Waterloo, Waterloo, ON, N2L 3G1, Canada}
\affiliation{Institute for Quantum Computing, University of Waterloo, Waterloo, ON, N2L 3G1, Canada}

\author{J.B. Kycia}
\affiliation{Department of Physics and Astronomy and Guelph-Waterloo Physics Institute, University of Waterloo, Waterloo, ON, N2L 3G1, Canada}
\affiliation{Institute for Quantum Computing, University of Waterloo, Waterloo, ON, N2L 3G1, Canada}

\author{B.D. Gaulin} 
\affiliation{Department of Physics and Astronomy, McMaster University,
Hamilton, Ontario, L8S 4M1, Canada}
\affiliation{Brockhouse Institute for Materials Research, McMaster University, Hamilton, Ontario, L8S 4M1, Canada}
\affiliation{Canadian Institute for Advanced Research, 180 Dundas St.\ W.,Toronto, Ontario, M5G 1Z8, Canada}


\bibliographystyle{prsty}

\begin{abstract}
Recent neutron scattering and specific heat studies on the pyrochlore Yb$_{2}$Ti$_{2}$O$_{7}$ have revealed variations in its magnetic behavior below 265mK.  In the best samples, a sharp anomaly in the specific heat is observed at T=265mK.  Other samples, especially single crystals, have broad features in the specific heat which vary in sharpness and temperature depending on the sample, indicating that the magnetic ground state may be qualitatively different in such samples. We performed detailed comparisons of the chemical structure of a pulverised single crystal of Yb$_{2}$Ti$_{2}$O$_{7}$, grown by the floating zone technique, to a sintered powder sample of Yb$_{2}$Ti$_{2}$O$_{7}$.  Rietveld refinements of neutron powder diffraction data on these samples reveal that the crushed single crystal is best described as a ``stuffed'' pyrochlore, Yb$_{2}$(Ti$_{2-x}$Yb$_{x}$)O$_{7-x/2}$ with $x$ = 0.046(4), despite perfectly stoichiometric starting material.   Substituting magnetic Yb$^{3+}$ on the non-magnetic Ti$^{4+}$ sublattice would introduce random exchange bonds and local lattice deformations.  These are expected to be the mechanism leading to of the variation of the delicate magnetic ground state of Yb$_{2}$Ti$_{2}$O$_{7}$. Determination of the cubic cell length, $a$, could be useful as a method for characterizing the stoichiometry of non-pulverised single crystals at room temperature.
\end{abstract}


\maketitle


\section{\label{sec:level1}Introduction}

The rare-earth titanate series of compounds, $R_{2}$Ti$_{2}$O$_{7}$, has been essential in furthering the understanding of geometrically frustrated magnetism over the past two decades.  Compounds in this series, in which $R$ stands for a trivalent rare earth (RE) ions, form perfect magnetic pyrochlore sublattices.  The pyrochlore lattice lends itself to studies of geometric frustration since it consists of corner-sharing tetrahedra.  The geometrical constraints imposed by this tetrahedral array frustrate near-neighbor magnetic interactions, coaxing rich behavior from even the simplest models for the spin Hamiltonian.\cite{lacroix2011introduction}  The rare earth titanates are realizations of this magnetic lattice where one has several choices of RE ions available, from Sm to Lu.\cite{subramanian1983oxide}  This choice introduces sufficient variability of the strongest terms in the spin Hamiltonians (such as single-ion anisotropy, dipolar interactions, and sign of exchange interactions) that a broad range of exotic magnetic behaviour has been observed across the series.\cite{gardner2010magnetic}  Furthermore, the relatively straightforward synthesis of large single crystals of these compounds, amenable to directionally dependent studies such as neutron scattering, has made them invaluable to the experimental side of the field.     

The synthesis of large single crystals of many compounds in this series can be accomplished by the optical floating zone (OFZ) method. \cite{dabkowska2010crystal, BalakrishnanFZ, KoohpayehFZoxide, prabhakaran2011crystal}  Aside from producing very large crystals (cylindrical in shape, with lengths of about 3cm and diameters of 0.5cm), this method is desirable since it is expected to keep impurities in the growths to a minimum by avoiding the use of a crucible.  Further, for congruently melting compounds such as the $R_{2}$Ti$_{2}$O$_{7}$ series, the chemical composition of the crystal is expected to stay constant, and identical to the starting material, over the length of the growth.  Despite these benefits,  recent studies have shown that there is some variability in the magnetic properties of single crystals of the same $R_{2}$Ti$_{2}$O$_{7}$ compound produced by this method.  This has been observed in Tb$_{2}$Ti$_{2}$O$_{7}$, where the magnetic specific heat was shown to vary from sample to sample, particularly in the single crystals.\cite{takatsu2012, yaouanc2011exotic, chapuisThesis}  Since Tb$_{2}$Ti$_{2}$O$_{7}$ is known for its lack of long range magnetic ordering down to 50mK, maintaining a disordered ground state with cooperatively fluctuating magnetic moments,\cite{gardner_spinliq} the effect of this sample dependence is not manifested in the ground state properties in an immediately obvious way, though a small fraction of spins appear to ``freeze'' in some samples. \cite{Gardner2003, luo2001low, lhotel2012low}  No work has been published that investigates the cause of this sample dependence in the single crystals, though a powder neutron diffraction study of a sintered powder found exact stoichiometry to within their experimental error of 2\%.\cite{Han2003}  We can assume that the sample dependence of the magnetic properties is the result of structural defects of some type which are so far uncharacterized in the single crystals, and are therefore uncontrolled.  

More recently, the same issue of sample dependence has arisen in Yb$_{2}$Ti$_{2}$O$_{7}$, but with much more dramatic consequences.   Yb$_{2}$Ti$_{2}$O$_{7}$ has long been an enigma when it comes to its magnetic ground state, which has been variously, and incompatibly, described as being long-range ordered in a collinear ferromagnetic (FM) state, \cite{yasui, chang2011higgs} having non-collinear long range FM fluctuations,\cite{gardner2004polarized} or having short-range correlated fluctuations.\cite{hodgesfluc, ross2009, ross2011dimensional}  The specific heat of Yb$_{2}$Ti$_{2}$O$_{7}$ is particularly revealing.  Some samples show sharp anomalies, with the sharpest and highest temperature anomaly observed so far being at 265mK in a powder sample prepared at McMaster University. \cite{ross2011dimensional}  Other powder samples show slightly broader anomalies at lower temperatures; 250mK\cite{dalmas2006studies, Yaouanc2011} and 214mK.\cite{blote}  The single crystal samples grown by the OFZ method in several laboratories are even more variable, showing very broad humps instead of sharp anomalies ,\cite{ross2011dimensional, Yaouanc2011, chang2011higgs} and in some cases show a mixture of both broad humps and sharp peaks.\cite{ross2011dimensional}  The loss of the extremely sharp specific heat anomaly, which signals a magnetic phase transition, likely means that some of the samples do not reach the ground state that would occur in a nominally ``perfect'' sample.   

Initial work on an effective Hamiltonian to describe Yb$_{2}$Ti$_{2}$O$_{7}$ was performed by Thompson \emph{et al}, who extracted four independent exchange parameters from fits to the single crystal diffuse magnetic neutron scattering above the transition.\cite{thompson2011rods}  While these parameters describe the local susceptibility of the moments well,\cite{thompson2011local} they do not reproduce the spin wave excitations that were revealed in a neutron scattering study of single crystal Yb$_{2}$Ti$_{2}$O$_{7}$ in the presence of a high magnetic field applied along the [110] direction.\cite{ross2009} These spin wave excitations, which present stronger constraints on the effective Hamiltonian, have themselves been accurately modeled by a different set of anisotropic exchange parameters.\cite{ross2011quantum}  These parameters revealed spin-ice type exchange, i.e. FM exchange between the local $\langle$111$\rangle$ components of the spins at the Yb$^{3+}$ sites, in addition to significant quantum fluctuations.  Yb$_{2}$Ti$_{2}$O$_{7}$ can thus be thought of as a ``quantum spin ice'' material, an analog to the classical spin ice state found in Ho$_2$Ti$_2$O$_7$ and Dy$_2$Ti$_2$O$_7$\cite{den2000dipolar} but with quantum dynamics allowing tunnelling between the many allowed spin ice ground states.\cite{ross2011quantum, shannon2012quantum, applegate2012yb2ti2o7, benton2012seeing} The precise location of Yb$_{2}$Ti$_{2}$O$_{7}$ on the four-exchange-parameter phase diagram of the S=1/2 anisotropic exchange model for the pyrochlore lattice is still unknown, but it appears to be close to the predicted Coulombic Quantum Spin Liquid (QSL) phase that is expected to remain disordered and support exotic emergent excitations such as magnetic monopoles.\cite{savary2011}  Another likely possibility is a simple FM ordered phase which is nearby in the phase diagram.\cite{savary2011}  This proximity to the phase boundary may mean that small perturbations to the exchange parameters, even locally, could induce a significant change in its ground state. 

The body of work on Yb$_{2}$Ti$_{2}$O$_{7}$ contains the following information on the sample dependence and the nature of the ground state:
\begin{itemize}
\item Powder samples tend to have a single specific heat anomaly, variable in both sharpness and temperature, reported to occur between 214mK and 265mK.\cite{blote, dalmas2006studies, Yaouanc2011, ross2011dimensional}
\item Single crystal samples tend to have broad humps in the specific heat. \cite{Yaouanc2011, ross2011dimensional}
\item A powder sample with a fairly sharp specific heat anomaly at T=250mK \cite{dalmas2006studies} shows a first-order drop in spin flucutation frequency, but remains dynamic below this transition and does not show signs for long-range order.\cite{hodgesfluc}
\item A single crystal sample with both a broad hump in the specific heat at 200mK and a sharp anomaly at 265mK shows a dynamic, short-range three-dimensionally correlated ground state that emerges from a short range two-dimensionally correlated state for T$\geq$400mK.\cite{ross2011dimensional}
\item One crystal has been reported to have a single, somewhat-sharp specific heat anomaly at 240mK.  This crystal shows some signs of a simple, almost-collinear FM ground state that emerges from a two-dimensionally correlated state that exists above the transition temperature. \cite{yasui, chang2011higgs} 
\end{itemize}

\begin{figure}[!htb]  
\centering
\includegraphics[ width=8.5cm]{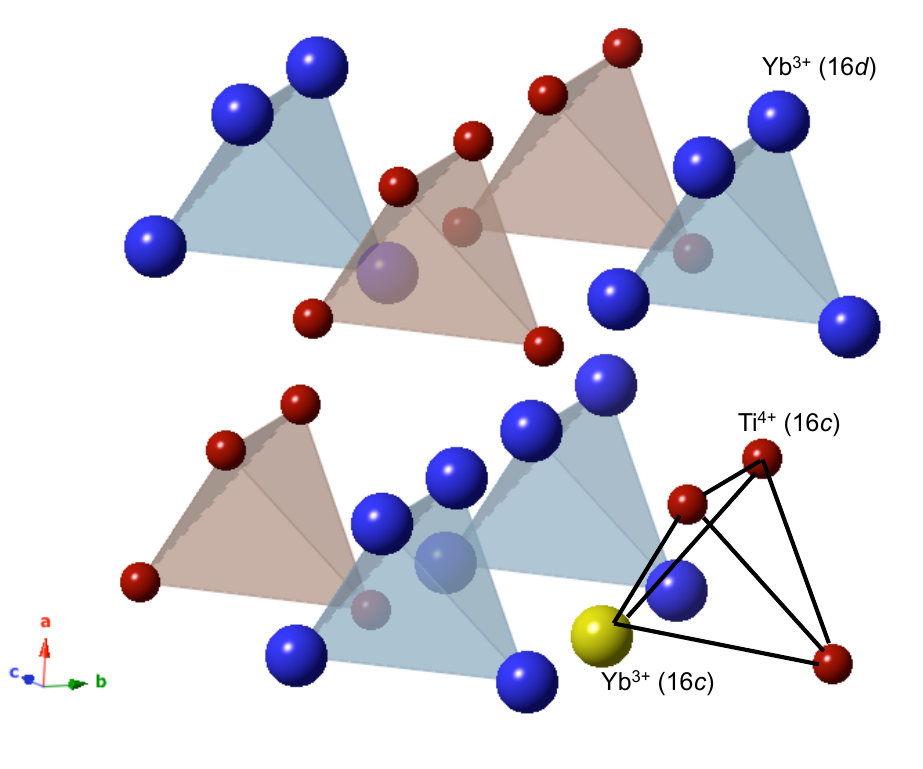}
\caption{ The pyrochlore lattice, showing only cation positions.  The stuffed pyrochlores include some RE cations (magnetic Yb$^{3+}$) on the transition metal (non-magnetic Ti$^{4+}$) site.  Here, this extra rare earth is shown in yellow.  This introduces extra near-neighbor exchange at some RE sites.}
\label{fig:fig1}
\end{figure}

Because the single crystal samples show variable, broad, and lower temperature specific heat features, one is lead to suspect that structural disorder is present in those samples to a higher degree than in the powders.  It is quite likely that these structural defects are small in number, since the growth conditions are well-known and repeatable, and investigation of the structures reveals perfectly ordered and stoichiometric compounds at first glance.\cite{ross2011dimensional}  Nevertheless, even small numbers of defects have the potential to cause significant changes in frustrated systems by disrupting the inherent delicate balance of energy scales present therein.  This may be particularly true for Yb$_{2}$Ti$_{2}$O$_{7}$ since it appears to be near a boundary in its exchange interaction phase diagram, as discussed above.  Therefore, we must look more closely at the structures of the single crystals, and at the possibility of subtle structural defects, to try to understand what implications they have for the magnetism in this material.  In the end, our goal should be to be able to characterize the level of such defects, which will allow us to systematically study their effects, since we must accept that real materials are never ideal.

With this goal in mind, we performed powder neutron diffraction experiments on two samples of Yb$_{2}$Ti$_{2}$O$_{7}$.  One sample is a sintered powder prepared identically to the starting material used in the OFZ growth of the single crystal.  Another powder sample prepared in this way has been shown to have the sharpest and highest temperature specific heat anomaly reported in the literature to date (this is reproduced in Fig.~\ref{fig:fig6}). \cite{ross2011dimensional}  The second sample is a single crystal grown by the OFZ method and then crushed (see Section \ref{sec:method} for details of the growth conditions).   We find, using Rietveld refinements of the neutron powder diffraction data, that the crushed single crystal is accurately modeled by full occupation of Yb$^{3+}$ ions on the Yb$^{3+}$ site, and by a 2.3\% substitution of Yb$^{3+}$ ions on the Ti$^{4+}$ sites.  

It is well-known that the $R_{2}$Ti$_{2}$O$_{7}$ pyrochlore materials can accommodate extra (magnetic) $R^{3+}$ ions on the (non-magnetic) Ti$^{4+}$ sites, leading to what is known as a ``stuffed'' pyrochlore structure. \cite{lau2006stuffed, Lau2007long}  The Ti sites themselves form a pyrochlore lattice which is interpenetrating with the $R$ sublattice (see Figure \ref{fig:fig1}), and the nearest neighbor (n.~n.) distance between Ti and $R$ sublattices is equal to the n.~n. distance between $R$ atoms; approximately 3.6 \AA.  

Stuffing introduces an extra n.n. bond for some RE ions which has different exchange pathway due to different oxygen environment, and thus may be of different strength or even sign.  This spatially random addition of a small number of exchange interactions, combined with the frustration already inherent in the system, could conceivably lead to a spin glass state.  This is consistent with the observation of broad humps in the magnetic specific heat in some of the single crystal samples of Yb$_{2}$Ti$_{2}$O$_{7}$, \cite{ross2011dimensional, Yaouanc2011} the formation of short range correlations with reduced fluctuation time scales,\cite{hodgesfluc, ross2011dimensional} and observed long-relaxation times \cite{yasui} with multiple timescales of relaxation,\cite{hodgesfluc} though no frequency-dependent ac susceptibility is reported at these low temperatures which could conclusively identify a spin glass transition.   We note that the frustrated garnet system Gd$_{3}$Ga$_{5}$O$_{12}$ (GGG) is also naturally and unavoidably ``stuffed'' with 1-2\% excess Gd on the Ga sites during crystal growth. ÊThis may also be the cause of the low-temperature unconventional glassiness that is observed in that system.\cite{schiffer1995frustration, quilliam2010juxtaposition}

There are reports of the magnetic properties of some stuffed rare-earth titanates with relatively high stuffing levels.  The stuffed spin ices,  Ho$_{2}$(Ti$_{2-x}$Ho$_{x}$)O$_{7-x/2}$ and Dy$_{2}$(Ti$_{2-x}$Dy$_{x}$)O$_{7-x/2}$ with $x=0.3$ (i.e. 15\% stuffing), show marked changes in spin dynamics compared to the unstuffed compounds, including the introduction of multiple timescales for relaxation, \cite{gardner2011slow, Ehlers2008} and slower low temperature dynamics in the Dy compound \cite{gardner2011slow}, yet more persistent dynamics in the Ho compound.\cite{zhou2007origin}  There is evidence for a change in n.n. exchange interactions, from ferromagnetic (FM) in the unstuffed spin ices, to anti-ferromagnetic (AFM) in the stuffed spin ices with $x=0.3$.\cite{zhou2007origin, Ehlers2008, gardner2011slow}   

In the Tb-based pyrochlores, a different type of disorder on the transition metal site has been investigated.  Tb$_2$Sn$_2$O$_7$ is known to reach an unusual ground state below T$_{c}$ = 850mK, consisting of both short- and long-range spin ordering,\cite{rule2007polarized, mirebeau2005ordered} while maintaining spin dynamics to the lowest measurable temperatures.\cite{rule2007polarized, mirebeau2005ordered, rule2009neutron} This is in  contrast to the isostructural compound Tb$_{2}$Ti$_{2}$O$_{7}$, which shows no signs of long range order at any measurable temperature.   Dahlberg \emph{et al} have shown that 5\% substitution of Ti for Sn in Tb$_2$Sn$_2$O$_7$ completely removes the transition to long range magnetic order.\cite{dahlberg2011low}  This example shows that the rare-earth pyrochlores can be very sensitive to small concentrations of defects on the non-magnetic sublattice, even when these substitutions are themselves non-magnetic in nature.

\begin{figure}[!htb]  
\centering
\includegraphics[ width=8.5cm]{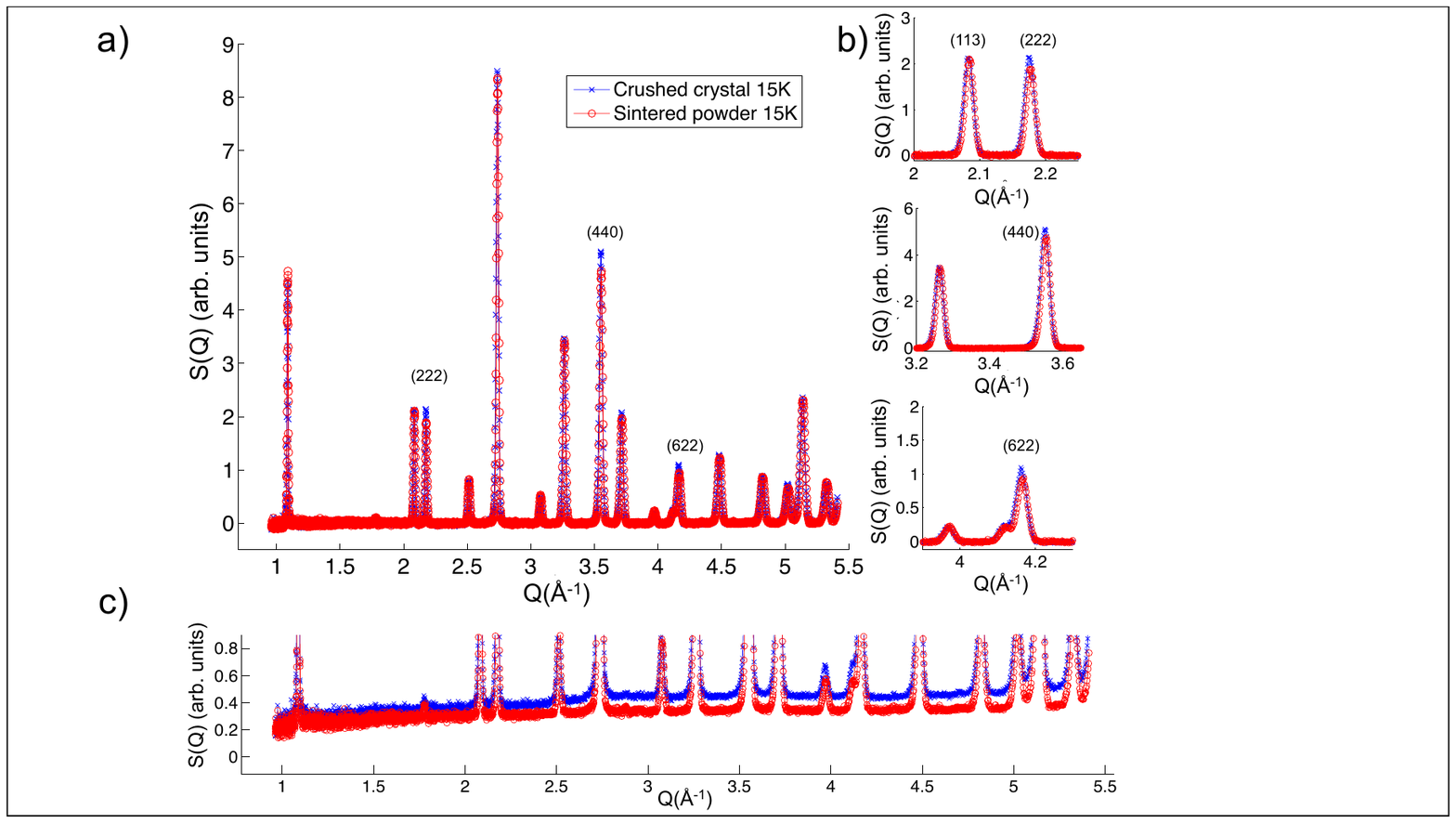}
\caption{ a) Powder neutron diffraction patterns at T=15K, obtained from a crushed single crystal and sintered powder.  The backgrounds, which were fit to 4th order polynomials, have been subtracted to compare peak heights.  Relative increases in $S(Q)$ are evident at the (222), (440) and (622) positions for the crushed crystal.  These Bragg peaks are shown in more detail in b).  There is an overall shift of the peak centers due to differences in lattice parameters, discussed in the text.  c) The raw data at T=15K without a background subtraction, focusing on the diffuse background.  The crushed crystal exhibits increased diffuse scattering. }
\label{fig:fig2}
\end{figure}

The structure of several stuffed rare earth titanates has been studied in detail by Lau \emph{et al}, \cite{lau2006stuffed, Lau2007long} who found that for lightly stuffed compounds, one observes on average a pyrohclore structure.  At higher levels of stuffing ($x=0.3$) for small RE ions such as Yb$^{3+}$, one observes a transition to defect fluorite structure with short range correlated pyrochlore superstructure.\cite{lau2006stuffed}  The lattice spacing of the cubic unit cell increases linearly with stuffing level, and has been characterized for several stuffed rare earth titanates including Yb$_{2}$(Ti$_{2-x}$Yb$_{x}$)O$_{7-x/2}$.\cite{lau2006stuffed}   We have drawn on this previous work to identify signatures in our data which indicate that the stuffed model is appropriate to the crushed crystal only, and quantify it using Rietveld refinements.  We further suggest that measurement of the room temperature lattice parameter is a viable method to ascertain the structural quality of single crystals of Yb$_2$Ti$_2$O$_7$ before performing further characterizations, such as low temperature specific heat.


\section{\label{sec:method}Experimental Method}

%

\subsection{Materials preparation}

Two samples of Yb$_2$Ti$_2$O$_7$ were studied by neutron powder diffraction.  One was prepared by solid state reaction between pressed powders of Yb$_2$O$_{3}$ and TiO$_{2}$, sintered at 1200$^{\circ}$C for 24 hours with warming and cooling rates of 100$^{\circ}$C/h.   A pressed powder prepared in this manner was also used as the starting material for the single crystal growth by the OFZ method.  The principles of this method are described elsewhere. \cite{dabkowska2010crystal, BalakrishnanFZ, KoohpayehFZoxide, Gardner_growth}  The growth rate was 6mm/h, with a counter-rotation of the feed and seed rods of 30 rpm which encouraged thorough mixing of the molten zone.  After the growth, no evaporated material was visually observed on the quartz tube that encloses the OFZ growth.  The growth was performed under 4atm of oxygen pressure.  Oxygen overpressure has previously been found to minimize oxygen non-stoichiometry in rare-earth titanates grown by the floating zone method. \cite{prabhakaran2011crystal}  

Both types of samples were then pulverised using a Pulverisette 2 mortar grinder, for 20 minutes per sample, to obtain loose powder samples.  They are henceforth referred to as the ``sintered powder'' (SP), weighing 12.587g,  and ``crushed crystal''(CC), weighing 10.294g.  

A second single crystal was prepared the same manner as described above, and a single slice of the crystal was annealed in flowing oxygen gas for 10 days, at 1050$^{\circ}$C, with warming/cooling rates of 100$^{\circ}$C/h.  This sample was not pulverised and was used in specific heat measurements.  Its lattice spacing was determined using a Huber four-circle x-ray diffractometer with an molybdenum anode.   A second slice of of the same crystal was not annealed in order to compare its specific heat to the annealed piece.

\subsection{Neutron powder diffraction}
A neutron powder diffraction experiment was performed using the NPDF instrument at the Lujan Neutron Scattering Center at the Los Alamos National Laboratory.  This instrument is located on a 32m flight path from the spallation target at the Lujan Center and has four banks of detectors, allowing the collection of high-resolution $S(Q)$ data ($\Delta Q/Q \sim$ 0.7\%) over a $Q$ range of 0.8 to 50\AA$^{-1}$.   The lowest temperature used in our data collection was 15K, which is well above both the Curie-Weiss temperature, $\Theta_{CW} \sim 800$mK, and the temperature of the observed magnetic phase transition, $T_{c}$ = 265mK; hence, all information in the measured $S(Q)$ is expected to arise from nuclear (chemical) structure rather than from magnetic correlations.

\subsection{Specific heat}
The specific heat measurements were performed at the University of Waterloo using the quasi-adiabatic method (see Ref. \onlinecite{quilliam2007specific} for details) with a 1 k$\Omega$ RuO$_2$ thermometer and 10 k$\Omega$ heater mounted directly on the thermally isolated sample.  A weak thermal link to the mixing chamber of a dilution refrigerator was made using Pt-W (92\% Pt, 8\% W) wire for the annealed and non-annealed single crystal pieces (354.1mg and 361mg, respectively) and powder sample (24.37 mg).    The time constant of relaxation provided by the weak link was several hours, much longer than the internal relaxation time of the samples, minimizing thermal gradients and ensuring that the sample cooled slowly into an equilibrium state.  The addenda contributed less than 0.1\% to the specific heat of the system.  The data from the powder sample was previously published in Ref. \onlinecite{ross2011dimensional}.

\begin{figure}[!htb]  
\centering
\includegraphics[ width=\columnwidth]{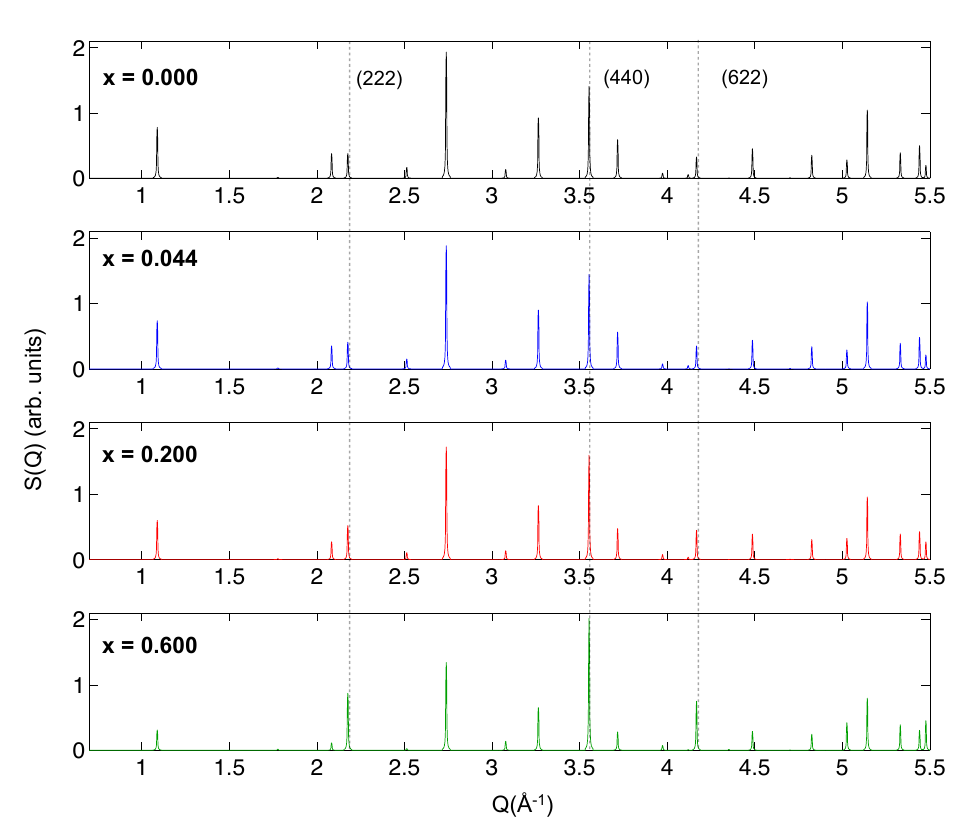}
\caption{ Calculated diffraction patterns for increasing levels of stuffing ($x$ = 0.000, 0.044, 0.200, and 0.600 from top to bottom, in Yb$_{2}$(Ti$_{2-x}$Yb$_{x}$)O$_{7-x/2}$).  The most obvious change, particularly at lower stuffing levels, is the ``switching'' of relative intensity at (222) and its neighboring peak, (113).   Other significant changes are the intensities of (622) and (440), which both increase for increased stuffing levels.   }
 \label{fig:fig3}
\end{figure}

\begin{figure*}[!htb]  
\centering
\includegraphics[ width=\textwidth]{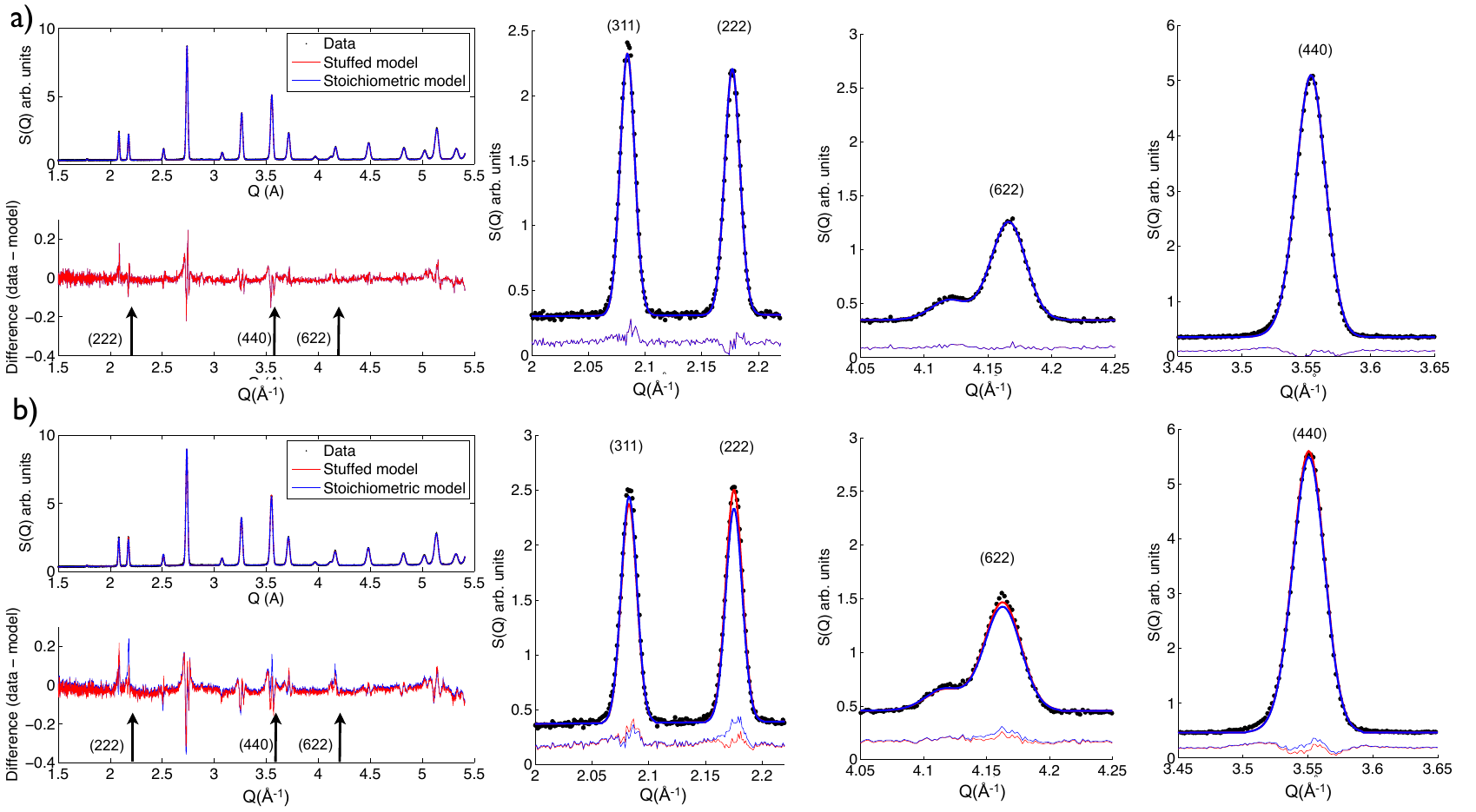}
\caption{ A comparison of two models used in the refinement of a) the powder sample and b) crushed crystal sample.  This figure shows data at T=15K only, as data at all measured temperatures are similar.  The stoichiometric model, corresponding to Yb$_{2}$Ti$_{2}$O$_{7}$, describes the powder sample well.  For the crushed crystal sample, improvements in modelling several peak heights are obtained using a ``stuffed'' model corresponding to Yb$_{2}$(Ti$_{2-x}$Yb$_{x}$)O$_{7-x/2}$, with $x$ = 0.046(4) (2.3\% substitution of Yb$^{3+}$ for Ti$^{4+}$).}
 \label{fig:fig4}
\end{figure*}

\begin{figure*}[!htb]  
\centering
\includegraphics[ width=\textwidth]{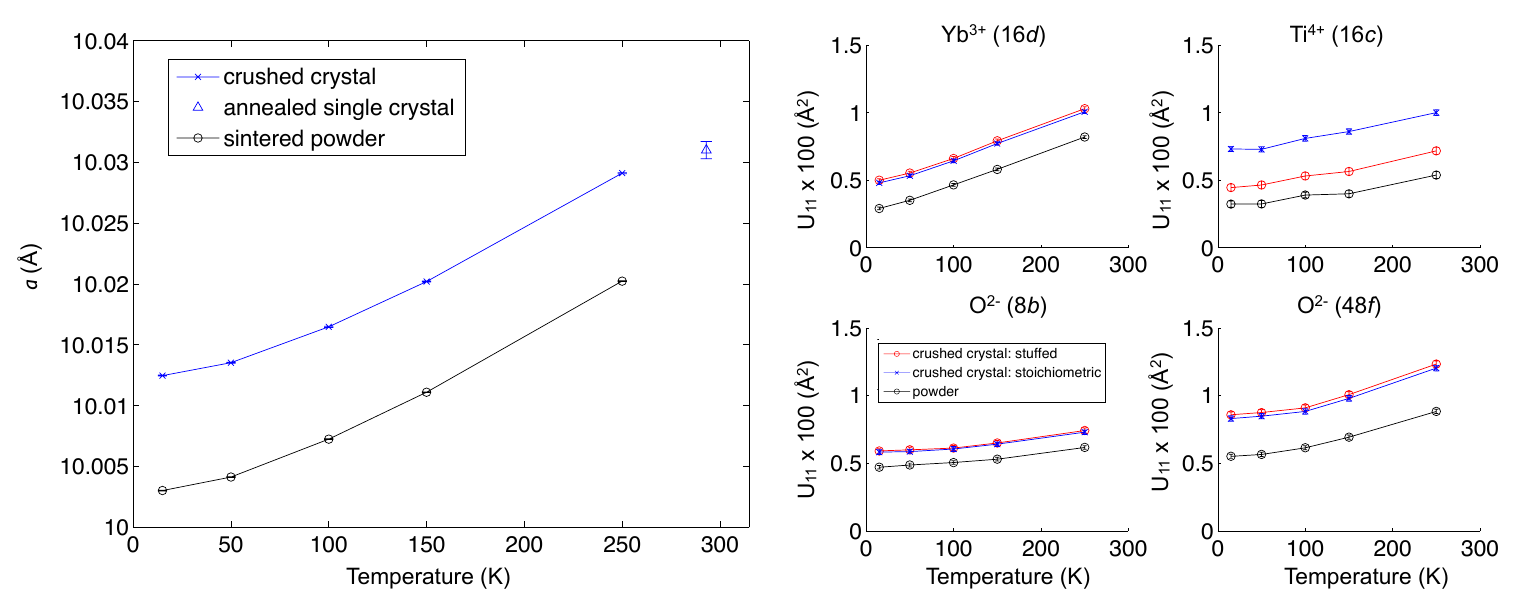}
\caption{ Left: Refined cubic lattice parameter, $a$, for the crushed crystal and sintered powder samples.  The crushed crystal shows a larger lattice parameter when compared to the sintered powder, as expected for a stuffed pyrochlore.  A non-pulverised single crystal sample post-annealed in oxygen gas shows a lattice parameter consistent with the crushed crystal sample.  The specific heat of the annealed crystal is shown in Fig. \ref{fig:fig6}.  Right: Selected anisotropic displacement parameters (ADPs) for each lattice site, refined in both the stoichiometric and stuffed models for the crushed crystal, and the stoichiometric model for the sintered powder.  Overall, the ADPs are found to be larger in the crushed crystal than in the sintered powder.  The Ti$^{4+}$ ADPs are the only ones that are affected by choice of model, to within the errorbars.  They are significantly reduced in the stuffed model, making the difference in Ti$^{4+}$ ADPs between the crushed crystal and sintered powder consistent with the other sites.  The ADPs for the 48$f$ O$^{2-}$ sites remain relatively high for both models, indicating the presence of true static displacements of the oxygen ions surrounding the stuffed 16$c$ sites.}
 \label{fig:fig5}
\end{figure*}

\begin{figure}[!htb]  
\centering
\includegraphics[ width=\columnwidth]{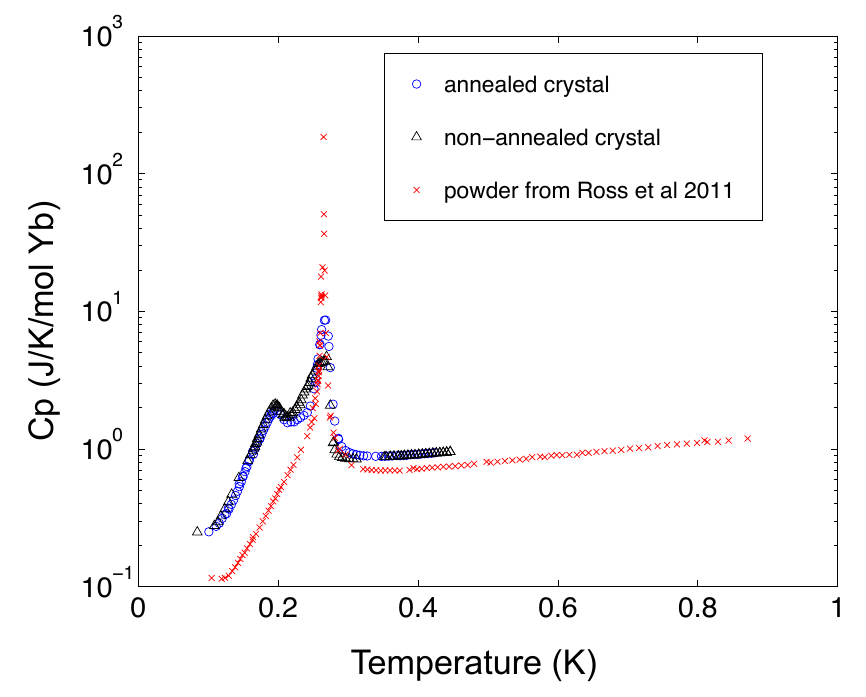}
\caption{ Specific heat of a single crystal annealed in oxygen gas for 10 days (blue circles), compared to a non-annealed slice of the same crystal (black triangles), as well as the powder sample from Ref. \onlinecite{ross2011dimensional} (red x's).}
 \label{fig:fig6}
\end{figure}


\section{\label{sec:level1}Results}


\subsection{Rietveld refinements}

 Neutron powder diffraction data were collected over a temperature range from 15 to 250K for the sintered powder and crushed crystal samples.  Representative data at T=15K are compared in Figure \ref{fig:fig2}, which shows S$(Q)$ as measured by the lowest angle bank of detectors in the NPDF instrument.  Rietveld structural refinements were carried out using EXPGUI,\cite{toby2001expgui} a graphical interface for the GSAS software package.\cite{larson2000generalized}   Note that in all refinements presented herein, all four banks of detectors are refined simultaneously, but only one is shown in the figures for clarity.   All data sets were refined within the Fd$\overline{3}$m space group, which is appropriate for the rare-earth titanate pyrochlore materials.  The positions of the ions and their Wyckoff positions are listed in Table \ref{tab:1}.  In the fully stoichiometric compound, both the magnetic Yb$^{3+}$ ions and the Ti$^{4+}$ ions form interpenetrating arrays of corner sharing tetrahedra (Fig. \ref{fig:fig1}), which we call sublattices A and B.  The Yb$^{3+}$ ions, normally confined to sublattice A, are 8-fold coordinated, with six of the O$^{2-}$ ions at the 48$f$ positions and two at the 8$b$ positions.  The Ti$^{4+}$ ions are 6-fold coordinated by O$^{2-}$ ions at the 48$f$ positions.  The only free positional parameter in this structure belongs to the 48$f$ oxygen ions. 

\begin{table}[t]
\begin{center}
\begin{tabular}{l | c c c c c}
\hline
\multicolumn{1}{l|}{Atom}&\multicolumn{1}{c|}{Wyckoff position}&\multicolumn{3}{c|}{ fractional coordinates}&\multicolumn{1}{c}{Ideal occ.}\\\hline
Yb1  & 16$d$ & 0.500 & 0.500 & 0.500 & 1\\
Yb2  & 16$c$ & 0.000 & 0.000 & 0.000 & 0\\
Ti  & 16$c$ & 0.000 & 0.000 & 0.000 & 1\\
O1  & 8$b$ & 0.375 & 0.375 & 0.375 & 1\\
O2  &  48$f$ & $x$ & 0.125 & 0.125 & 1\\
\hline
\hline
\end{tabular}
\caption{\label{tab:1} Crystallographic sites used in the refinements.  Their Wyckoff positions and fractional coordinates are given within the 2nd origin choice of the  Fd$\overline{3}$m space group.}
\end{center}
\vspace{-0.6cm}
\end{table}

In our refinements, the Atomic Displacement Parameters (ADPs) are allowed to take their full anisotropic forms, which are contstrained only by the point symmetry of the site in question.  These constraints are laid out in Ref. \onlinecite{greedan2009local}.  The ADPs represent the mean-squared displacement of the ions from their average positions.  Thus, one expects them to increase as temperature increases, in order to account for thermal motion.  The ADPs also include any \emph{static} displacements, such as those due to local distortions from the ideal structure, which are averaged over all unit cells.  For a stuffed pyrochlore, we should expect the ADPs for all ions to increase mildly due to local strain caused by the inclusion of the (relatively large) Yb ion at the Ti site, and the resulting oxygen vacancies required to maintain charge neutrality.  We do observe increased ADPs for the crushed crystal for all refined models.

For both samples the structure refines quite well within the purely stoichiometric model, with combined $R_{wp}$ (i.e. the ``weighted profile $R$ factor'', a measure of the goodness of the fit which is ideally zero\cite{toby2006r}) values for all detector banks between 4.0 and 4.8\% (see Table \ref{tab:2}).  The crushed crystal refined in the stoichiometric model does give poorer fits, on the 0.4\% level, compared to the sintered powder.  Examining the raw data reveals subtle differences between the two types of samples, which indicate that the condition for exact stoichiometry in the crushed crystal should be relaxed.  Examples of the raw neutron powder diffraction data are shown in Fig. \ref{fig:fig2}, which have been normalized to the incident spectrum and have diffuse backgrounds subtracted in panels a) and b).  Figure \ref{fig:fig2} a) shows that the relative intensities of some Bragg reflections differ between the two types of samples, most noticeably the (222), (440) and (622) peaks, while b) shows these peaks in more detail.  Here, one can also observe that there is a difference in lattice spacing, $a$, which is found to be on average 0.08\% larger for the crushed crystal (see Table \ref{tab:2}). Though it is not obvious from the figure, the widths of the Bragg peaks are also larger in the crushed crystal, likely indicating some amount of lattice strain.     Finally, the diffuse background, shown in Fig. \ref{fig:fig2}c), is more appreciable in the crushed crystal.  Diffuse scattering, which arises from the inclusion of phonon scattering in the diffraction pattern as well as uncorrelated or extremely short range correlated atomic positions, can be viewed as the complement of larger ADPs; the ``missing'' intensity characterized by the ADPs goes into this diffuse background.

The higher diffuse background indicates that some amount of structural disorder is present.  Meanwhile, the differences in relative intensity at certain Bragg positions, such as (222), (440) and (622), indicate an average difference in an ordered structure, which may be independent of whatever strain or positional disorder is present.  The structural difference leading to these changes in relative peak intensities should be quantifiable using Rietveld refinements of well-chosen structural models.  This is what we now focus on.


\begin{table*}[t]
\begin{center}
\begin{tabular}{l | c c c c | c c c}
\hline
\multicolumn{8}{c}{Sintered Powder}\\\hline
 \multicolumn{1}{c|}{}&\multicolumn{4}{c|}{Stuffed Model}&\multicolumn{3}{c}{Stoichiometric Model}\\\hline
T(K) &  Yb 16$c$ occupancy& $a$(\AA) & $x$ (O2) & R$_{wp}$ & $a$(\AA) & $x$ (O2) & R$_{wp}$ \\
\hline
15 & 0.000(1) & 10.00302(3) & 0.33146(3)& 0.0424& 10.00302(3) & 0.33147(3)& 0.0422\\

50 & 0.002(1) & 10.00413(3) & 0.33141(3) & 0.0444 & 10.00413(3) & 0.33143(3)& 0.0442\\

100 & 0.001(1) & 10.00724(3) & 0.33133(3) & 0.0409& 10.00724(3) & 0.33134(3)& 0.0407\\

150 & 0.002(1) & 10.01111(3)  & 0.33121(3) & 0.0415 & 10.01111(3) & 0.33122(3)& 0.0414\\

250 & 0.001(1) & 10.02024(3)  & 0.33103(3) & 0.0380 & 10.02024(3) & 0.33104(3)& 0.0379\\
\hline
\hline
\multicolumn{8}{c}{Crushed Crystal}\\\hline
\multicolumn{1}{c|}{}&\multicolumn{4}{c|}{Stuffed Model}&\multicolumn{3}{c}{Stoichiometric Model}\\\hline
T(K) &  Yb 16$c$ occupancy & $a$(\AA) & $x$ (O2) & R$_{wp}$ & $a$(\AA) & $x$ (O2) & R$_{wp}$ \\
\hline
15 & 0.023(2) & 10.01247(5) & 0.33180(4) & 0.0457 & 10.01247(5) & 0.33193(4)& 0.0460\\

50 & 0.023(2) & 10.01353(5) & 0.33179(4) & 0.0477 &  10.01353(5) & 0.33191(4) & 0.0472\\

100 & 0.026(2) & 10.01648(5) & 0.33168(5) & 0.0477 & 10.01648(5) & 0.33181(4) & 0.0480\\

150 & 0.023(2) & 10.02020(5) & 0.33156(4) & 0.0450 & 10.02020(5) & 0.33170(4) & 0.0453\\

250 & 0.026(2) & 10.02910(4) & 0.33139(4) & 0.0415 & 10.02910(4) & 0.33154(4) & 0.0419\\
\hline
\hline
\end{tabular}
\caption{\label{tab:2}Basic refined parameters for the sintered powder and crushed crystal sample, using the stuffed model and the stoichiometric model}
\end{center}
\vspace{-0.6cm}
\end{table*}

\subsubsection{Relative peak intensities}

Reports have been published on the structure of ``stuffed'' rare earth titanates, which have shown the evolution of the diffraction pattern as more rare-earth is placed on the titanium site.\cite{Lau2007long}  A calculation of these patterns is shown in Fig. \ref{fig:fig3} for Yb$_{2}$(Ti$_{2-x}$Yb$_{x}$)O$_{7-x/2}$.  The most obvious change at low levels of stuffing is the increase of (222) relative to its immediate neighbor, (113).  This is clearly observed in the raw data shown in Fig.~\ref{fig:fig2}.  Other expected changes coincide with differences that are readily observed in our data as well, such as the relative increase of the (440) and (622) peaks.  We are thus lead to investigate the role of stuffing in these samples.

We now present a comparison between the stoichiometric and stuffed model refinements for both samples at each measured temperature.  All of the refined parameters for each case are shown in Tables \ref{tab:2} and \ref{tab:3}.  Figure \ref{fig:fig4} shows the results of refinements in both models.  The curve in the lower left panels for these figures represents the difference between the data and the model (residuals).  In the case of the crushed crystal, larger differences are seen at the (222), (440) and (622) positions when the stoichiometric model is used (blue curve), compared to when the stuffed model is used (red curve).  These peaks and the difference curves are shown in more detail on the right hand panels of Fig. \ref{fig:fig4}, where the difference curves have been shifted up by a constant for clarity.  The stuffed model refines to $x$=0.046, or 2.3\% stuffing, for the crushed crystal.  Meanwhile, the powder sample refines to $x$=0.000(1), and no difference is observed in the stoichiometric vs. stuffed patterns (Fig.~\ref{fig:fig4}a).

\subsubsection{Lattice spacing}
The lattice spacing of the crushed crystal sample is larger than the sintered powder (see Fig. \ref{fig:fig5}).  This is generally consistent with stuffing in pyrochlores, as shown in Figure 3 of Ref. \onlinecite{lau2006stuffed}.  The increased cell size simply arises to accommodate the larger rare-earth ions in the titanium sites.  The work by Lau \emph{et al} in principle allows a correspondence between $x$ and the room temperature lattice spacing of Yb$_{2}$(Ti$_{2-x}$Yb$_{x}$)O$_{7-x/2}$.  Our maximum measured temperature was 250K, so that we must extrapolate to room temperature to compare to this study.  Quadratic fits of $a$ vs $T$ produce a room temperature lattice spacing for the powder sample that is well-below the minimum value reported by Lau \emph{et al}, which was 10.032 \AA\ at  $x$=0.  We find $a_{293K}$ = 10.025 \AA\ for the powder, and $a_{293K}$ = 10.034 \AA\ for the crushed crystal.  It is unclear as to why there is a discrepancy in the stoichiometric lattice spacing.  However, we note that if one assumes a systematic error in the lattice spacing determination, perhaps due to instrumental effects, the difference between the lattice spacing of the crushed crystal and the sintered powder (the latter taken to be stoichiometric) would coincide with $x$ = 0.032 according to Figure 3 of Ref \onlinecite{lau2006stuffed}.  This is in fairly good agreement with the refined value of $x$=0.046.
We also note that for a nominally stoichiometric Yb$_{2}$Ti$_{2}$O$_{7}$ powder sample reported on elsewhere,\cite{bramwell_mag} the room temperature lattice spacing was found to be 10.024(1) \AA.  This is in better agreement with the values found here for the sintered powder sample.  

We checked the lattice spacing of an additional single crystal of Yb$_2$Ti$_2$O$_7$, which has been characterized by low temperature specific heat.  This single crystal was post-annealed in flowing oxygen for 10 days, at 1050$^{\circ}$ C. The specific heat curve is are shown in Fig.~\ref{fig:fig6}, and compared to the powder sample from Ref. \onlinecite{ross2011dimensional}.  Consistent with other crystals of Yb$_2$Ti$_2$O$_7$, the specific heat curve of the annealed single crystal shows two relatively broad anomalies.  This probably rules out both lattice strain and oxygen deficiency as the causes of the sample dependence, both of which would be expected to be relieved by the heat treatment, since it does not significantly improve the sharpness and homogeniety of the speicifc heat anomaly in the crystals.  Furthermore, the lattice parameter for the annealed sample, $a = 10.0310(7)$, is consistent with that found for the CC sample studied here by neutron powder diffraction (Figure \ref{fig:fig5}).

\subsubsection{Atomic Displacement Parameters}
Aside from capturing the relative peak intensity changes in the powder diffraction pattern for the crushed crystal, the stuffed model also suppresses the anomalously high ADPs that are refined for the Ti site in the stoichiometric model.  The right hand side of Figure \ref{fig:fig5} shows one component of the ADPs on each site as a function of temperature, for both powder and crushed crystal samples.  A general increase in ADPs is expected for a more disordered sample, and this is born out by the small increase observed in ADPs at the Yb(16$d$) and O(8$b$) sites for both structural models refined for the crushed crystal. However, both the O(48$f$) the Ti(16$c$) sites see more dramatic increases in ADPs for the crushed crystal when the stoichiometric model is used.  In the case of the O(48f) site, this large ADP is consistent with a stuffed pyrochlore, since static displacements of the oxygen environment surrounding the B site are expected to arise from the valency mismatch of the Yb$^{3+}$ and Ti$^{4+}$ ions on the 16$c$ site.  We find that the large ADP at the Ti site, however, is alleviated by using the 2.3\% stuffed model.  In this case, the large ADP value found for the 16$c$ site when refined within the stoichiometric model is likely to be ``unphysical'', in the sense that it is compensating for problems in the assumed model.  In general, the ADP effectively accounts for a ``smearing'' of scattering weight around a particular position, which changes the effective scattering length at that site. This can occur because of temperature increase or static distortions (both are legitimate structural effects), or because the wrong effective scattering length is used in the model.  The latter appears to be the case here for the 16$c$ site.  Using the stuffed model instead, which places some ytterbium on the titanium site, brings the 16$c$ ADPs down to approximately the same levels as the other sites. This is a strong indication that the crushed crystal is actually a lightly stuffed pyrochlore.

\subsection{Susceptibility}

\begin{figure}[!htb]  
\centering
\includegraphics[ width=\columnwidth]{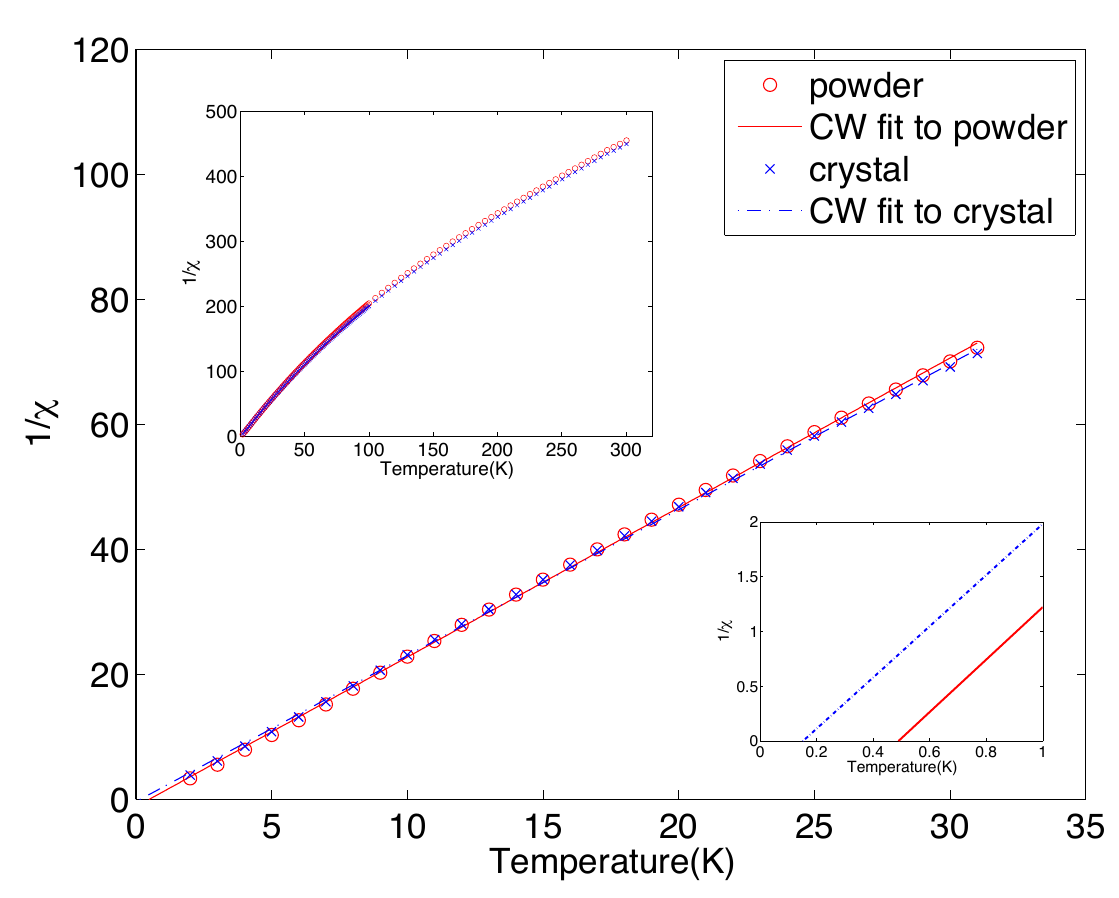}
\caption{ Inverse volume susceptibility (dimensionless) for both sintered powder and crushed crystal samples, with fits to a Curie-Weiss (CW) law ($\chi^{-1} =  \frac{T - \theta_{CW}}{C}$).  Top inset: inverse susceptibility over the full temperature range, 2K to 300K.  Bottom inset: extrapolation to 1/$\chi$=0 of the CW fits.}
 \label{fig:fig7}
\end{figure}

Inverse susceptibilty as a function of temperature, collected at a constant magnetic field of 0.2T, is plotted for both SP and CC samples in Fig. \ref{fig:fig7}.  The top inset shows the full temperature range over which data was collected.  The inverse susceptibility is clearly non-linear over the full temperature range, 2K to 300K; this has been noted already in the literature (Refs. \onlinecite{bramwell_mag} and \onlinecite{cao2009aniso}).  Population at the 10$\%$ level of the 680K crystal field excitation may be partially responsible for the non-linearity at high temperatures.  However, even the lower temperature behavior (below 30K) is slightly non-linear, which is known to be a result of the magnetic anisotropy present in Yb$_{2}$Ti$_{2}$O$_{7}$.\cite{cao2009aniso}  Fitting the data from 2K to 30K to a Curie-Weiss law without taking these anisotropies into account, we extract an effective moment of 3.14$\mu_{B}$ for the SP sample, and 3.43$\mu_{B}$ for the CC, consistent with an increased number of moments in the latter sample.  The Curie-Wiess temperatures were $\theta_{CW} = 152$mK for the CC sample, and $\theta_{CW} = 490$mK for the SP sample.   Taken at face value, this implies that the CC sample is slightly more anti-ferromagnetic in nature than the SP sample, though the average interaction is still ferromagnetic.  We comment that fitting from 2K to 10K, as was done in Refs.~\onlinecite{hodgescrysfield} and \onlinecite{cao2009aniso}, gives $\theta_{CW} = 733$mK for the SP sample, in fair agreement with those studies (which found 800mK and 750mK, respectively).   The inverse susceptibility for the CC sample fit over this smaller range gives $\theta_{CW} = 462$mK.

\section{\label{sec:level1}Discussion}

\begin{table*}[t]
\begin{center}
\begin{tabular}{l | c c c c c c c c}
\hline
\multicolumn{9}{c}{Sintered Powder (stoichiometric model)}\\\hline
  \multicolumn{1}{c|}{}&\multicolumn{8}{c}{ADPs in \AA$^{2}$}\\\hline
  \multicolumn{1}{c|}{T(K)}&\multicolumn{2}{c|}{Yb (16$d$) U$_{eff}$}&\multicolumn{2}{c|}{Ti (16$c$) U$_{eff}$}&\multicolumn{1}{c|}{O1 (8$b$) U$_{eff}$}&\multicolumn{3}{c}{O2 (48$f$) U$_{eff}$}\\\hline
250 & \multicolumn{2}{c}{0.0082(1)} &\multicolumn{2}{c}{0.0057(1)} & \multicolumn{1}{c}{0.0062(1)}& \multicolumn{3}{c}{0.0078(2)}\\
\hline
 &  U11(Yb 16$d$) & U12(Yb 16$d$)) & U11(Ti 16$c$) & U12(Ti 16$c$) & U11(O1 8$b$) & U11(O2 48$f$) & U22(O2 48$f$) & U23(O2 48$f$)\\
\hline
15 & 0.0029(1) & -0.00016(6)& 0.0033(1) & -0.00104(2)& 0.0047(1) & 0.0055(2) & 0.0051(1) & 0.0004(1)\\

50 & 0.0035(1)& -0.00038(6)& 0.0034(1)& -0.00105(2)& 0.0048(1)& 0.0056(2)& 0.0051(1)& 0.0005(1)\\

100 & 0.0046(1)& -0.00075(6)& 0.0040(1)& -0.00109(2)& 0.0050(1)& 0.0061(2)& 0.0055(1)& 0.0006(1)\\

150 & 0.0058(1)& -0.00116(6)& 0.0043(1)& -0.00120(2)& 0.0053(1)& 0.0069(2) & 0.0059(1) & 0.0008(1)\\

250 & 0.0082(1) & -0.00192(6)& 0.0058(1)& -0.00109(2)& 0.0062(1) & 0.0088(2)& 0.0072(1) & 0.0017(1)\\
\hline
\hline
\multicolumn{9}{c}{Crushed Crystal (stuffed model)}\\\hline
  \multicolumn{1}{c|}{}&\multicolumn{8}{c}{ADPs in \AA$^{2}$}\\\hline
   \multicolumn{1}{c|}{T(K)}&\multicolumn{2}{c|}{Yb (16$d$) U$_{eff}$}&\multicolumn{2}{c|}{Ti (16$c$) U$_{eff}$}&\multicolumn{1}{c|}{O1 (8$b$) U$_{eff}$}&\multicolumn{3}{c}{O2 (48$f$) U$_{eff}$}\\\hline
250 & \multicolumn{2}{c}{0.0103(1)} &\multicolumn{2}{c}{0.0072(2)} & \multicolumn{1}{c}{0.0074(2)}& \multicolumn{3}{c}{0.0107(2)}\\
\hline
 &  U11(Yb 16$d$) & U12(Yb 16$d$)) & U11(Ti 16$c$) & U12(Ti 16$c$) & U11(O1 8$b$) & U11(O2 48$f$) & U22(O2 48$f$) & U23(O2 48$f$)\\
\hline
15 & 0.0050(1) & -0.000990(7)& 0.0045(3) & -0.0013(3) & 0.0059(2) & 0.0086(2) & 0.0074(2) &  0.0007(1)\\

50 & 0.0055(1)& -0.00116(7)& 0.0046(3) & -0.0015(3) & 0.0060(2) & 0.0088(2)& 0.0075(2) & 0.0008(2) \\

100 & 0.0066(1) & -0.00155(8)& 0.0053(3) & -0.0017(3) & 0.0061(2) & 0.0091(2) & 0.0079(2) & 0.0008(2)\\

150 & 0.0079(1)& -0.00201(8) & 0.0056(3) & -0.0013(3) & 0.0065(2) & 0.0101(2)& 0.0084(2)& 0.0013(2) \\

250 & 0.0103(1) & -0.00269(6) & 0.0072(2) & -0.0015(3) & 0.0074(2) & 0.0123(2) & 0.0099(2)& 0.0020(2) \\
\hline
\hline
\multicolumn{9}{c}{Crushed Crystal (stoichiometric model)}\\\hline
  \multicolumn{1}{c|}{}&\multicolumn{8}{c}{ADPs in \AA$^{2}$}\\\hline
     \multicolumn{1}{c|}{T(K)}&\multicolumn{2}{c|}{Yb (16$d$) U$_{eff}$}&\multicolumn{2}{c|}{Ti (16$c$) U$_{eff}$}&\multicolumn{1}{c|}{O1 (8$b$) U$_{eff}$}&\multicolumn{3}{c}{O2 (48$f$) U$_{eff}$}\\\hline
250 & \multicolumn{2}{c}{0.0101(1)} &\multicolumn{2}{c}{0.0100(2)} & \multicolumn{1}{c}{0.0073(2)}& \multicolumn{3}{c}{0.0104(2)}\\
\hline
 &  U11(Yb 16$d$) & U12(Yb 16$d$)) & U11(Ti 16$c$) & U12(Ti 16$c$) & U11(O1 8$b$) & U11(O2 48$f$) & U22(O2 48$f$) & U23(O2 48$f$)\\
\hline
15 & 0.0048(1) & -0.00094(7)& 0.0073(2) & -0.0026(3)& 0.0058(2) & 0.0083(2)& 0.0072(2)& 0.0008(2)\\

50 & 0.0053(1) & -0.00112(8)& 0.0073(2)& -0.0027(3)& 0.0059(2) & 0.0085(2)& 0.0073(2) & 0.0008(2)\\

100 & 0.0064(1)& -0.00151(8)& 0.0081(2)& -0.0029(3)& 0.0061(2) & 0.0088(2)& 0.0077(2) & 0.0008(2)\\

150 & 0.0077(1)& -0.00197(8)& 0.0086(2) & -0.0028(3)& 0.0064(2) & 0.0098(2)& 0.0081(2)& 0.0013(2) \\

250 & 0.0100(1) & -0.00305(6)& 0.0100(2)& -0.0028(3)& 0.0073(2) & 0.0120(2) & 0.0096(2)& 0.0021(2)\\
\hline
\hline
\end{tabular}
\caption{\label{tab:3}Anisotropic Displacement Parameters (ADPs) for the sintered powder and crushed crystal.  Results from both the stoichiometric and stuffed models are shown for the crushed crystal.  The first row of each section shows the effective isotropic ADPs at 250K, for quick comparison between models.  Other rows show the full symmetry-allowed ADPs at all measured temperatures. See Supplemental Material at \textbf{[URL]} for crystallographic information files containing the refined structures at 250K.}
\end{center}
\vspace{-0.6cm}
\end{table*}

Naively, the most likely scenario for the disorder in oxides grown by the OFZ method would have seemed to be oxygen non-stoichiometry, a factor that can be restricted to some extent by controlling the pressure and type of gaseous atmosphere during the growth.  Oxygen non-stoichiometry has been observed in several oxides grown by the OFZ technique, including the rare-earth titanate series,\cite{prabhakaran2011crystal} as well as some high T$_c$ superconductors. \cite{KoohpayehFZoxide}  In the case of some high T$_c$ superconducting materials, the transition to superconductivity is well-known to be sensitive to oxygen non-stoichiometry.\cite{tokura1988broader, pham1992} In the rare-earth titanates, oxygen deficiency is known to manifest as a change in color and transparency of the crystals.\cite{balakrishnan1998single}  Prabahkaran \emph{et al} have shown that the magnetization of the spin ice Dy$_2$Ti$_2$O$_7$ depends on the oxygen content, which was controlled in that study by post-annealing the crystals in various gaseous atmospheres \cite{prabhakaran2011crystal}.  This treatment also caused a significant change in the color of the crystals in that study.   In contrast, we have found evidence that simple oxygen non-stoichiometry in Yb$_2$Ti$_2$O$_7$ is \emph{not} responsible for the variance in the magnetic properties; upon annealing a single crystal of Yb$_2$Ti$_2$O$_7$ in 1atm of flowing O$_2$ gas for 10 days, we find no change in color or transparency, and relatively little improvement in the specific heat anomaly, which remains bifurcated and broad (see Fig.~ \ref{fig:fig6}).

Instead of a simple scenario of oxygen non-stoichiometry, the results of the refinements for the two samples of Yb$_2$Ti$_2$O$_7$ studied here indicate that the OFZ growth of Yb$_2$Ti$_2$O$_7$ is somehow deficient in titanium atoms.  This is evidenced by the larger ADPs at the Ti site in the refinements of the CC structure, as well as changes to the relative Bragg peak intensities.  We note that, given a titantium deficiency, the substitution of the rare earth for titanium is more likely to occur in Yb$_{2}$Ti$_{2}$O$_{7}$ than in the other titanates, since Yb has one of the smallest ionic radii of the series.\cite{lau2006stuffed}  The reason for the Ti deficiency is unclear, however, since the starting materials are stoichiometric (i.e. prepared identically to the SP sample studied here), and Yb$_2$Ti$_2$O$_7$ is nominally congruently melting, so that the molten zone should have the same composition as the starting material.  It is possible that there is a slight evaporation of Ti during crystal growth.  Evaporation of volatile components of the molten zone can be partially controlled by performing growths in elevated gas pressure.\cite{KoohpayehFZoxide} The CC studied here was grown at 4atm of O$_{2}$ gas pressure, and so this effect should be somewhat suppressed.  For appreciable evaporation it is possible to observe the buildup of excess material on the quartz tube that encloses the growth.  This was not visually observed in the growth of Yb$_2$Ti$_2$O$_7$, although we note that a 2.3\% evaporation of Ti would amount to only 34.6mg of material on the quartz tube for a $\sim$10g growth.  This may be difficult to see or measure by weighing the quartz tube, especially since nothing prevents the vapourized material from condensing in the upper components of the OFZ furnace away from the quartz tube.  

We note that Chang \emph{et al} \cite{chang2011higgs} have had success in producing a single crystal of Yb$_2$Ti$_2$O$_7$ via the OFZ method that displays an improvement in the sharpness of the specific heat anomaly.  Their growth procedure differed from ours in the following ways: they sintered their starting oxides at 1050$^{\circ}$C (we used 1200$^{\circ}$C), performed the growth in 1atm of air (we used 4atm of O$_2$), and used a much slower growth speed (1.5mm/h instead of our 6mm/h).  Of these changes, one may suspect that the growth atmosphere would be the only one to affect the titanium deficiency if evaporation is indeed the cause, but it is not immediately clear why growing in air and at a lower pressure would improve this.  The lower growth speed represents a much more dramatic difference in the growth process, but also does not seem to be directly related to titanium stoichiometry.  We suggest that further investigations of crystals grown in different gaseous atmospheres and with different growth speeds should be carried out to find the ideal conditions for this growth. 

We have found that the stuffed model more accurately accounts for the observed $S(Q)$ in the CC than does the stoichiometric model.  However, there are other possibilities involving disorder on the Ti site which should also be considered, namely anti-site disorder and Ti vacancies.  
\begin{itemize}
\item{\textbf{Anti-site disorder:}}  Switching Yb and Ti ions would seem to be a more natural type of disorder, since it does not require the loss of titanium through the growth process.  However, refining the CC within this model produces essentially an ideal structure with -0.4\% anti-site mixing (the negative sign indicates an unphysical \emph{over}-occupation of the sites with the original ions), and thus no change from the stoichiometric model in any other parameter.

\item{\textbf{Titanium vacancies:}}  This model (charge-balanced by oxygen vacancies) gives to a 6\% titanium deficiency with an improved R$_{wp} = 4.09\%$, compared to the stuffed model which gives R$_{wp} = 4.19\%$.  Despite this apparently better fit, this would seem to be an unlikely scenario since it requires the loss of even more titanium, which is already difficult to understand at the 2\% level.  Furthermore, vacancies would create \emph{positive} chemical pressure and tend to reduce the lattice spacing.  We observe the opposite effect in the CC sample.  
\end{itemize}

It should also be noted that the type of structural disorder here does not account for the observed presence of some forbidden Bragg peaks in single crystal measurements of many rare-earth titanates.\cite{rule_tbtio,clancy2009revisiting,ruff} The (002) and (006) peaks, forbidden within the Fd$\bar{3}$m space group, have been observed using time-of-flight neutron scattering.  This technique is not susceptible to contamination from the higher order neutron wavelengths that otherwise could ``artificially'' produce a small amount of scattering at systematically absent positions.  The presence of these unexpected reflections indicates either a breaking of the translational symmetry of the space group\cite{curnoe2008distortion} or another measurement effect such as multiple-scattering.  We note that the translational symmetry of the space group is not expected to be broken in the model we consider here, since we have assumed that the stuffing occurs at random titanium sites.   These reflections in Tb$_2$Ti$_2$O$_7$ are seen to vanish in high magnetic fields,\cite{Ruff2010} and therefore may be related to another (magnetoelastic) distortion that has so far escaped direct detection.


At this point, it is difficult to make definitive statements about the effect of stuffing on the magnetism in Yb$_2$Ti$_2$O$_7$.  Our susceptibility measurements show a tendency for the Curie-Weiss constant to decrease with stuffing, indicative of increased anti-ferromagnetic coupling.  This is consistent with the observation of anti-ferromagnetic correlations in highly-stuffed spin ices.    One could expect the introduction of additional nearest neighbor bonds between Yb ions to dramatically effect the ground state if geometric frustration is key to its selection.  The determination of the exchange Hamiltonian for Yb$_2$Ti$_2$O$_7$ via high field inelastic neutron scattering has shown that spin-ice type exchange (ferromagnetic exchange along the local $\langle$111$\rangle$ directions on the magnetic pyrochlore sublattice) exists in Yb$_2$Ti$_2$O$_7$ and thus creates geometric frustration.  This, combined with quantum fluctuations, suppresses the transition to long range order by at least 1 order of magnitude.\cite{ross2011quantum}  Thus, one could expect that introducing randomly located exchange bonds (with different strengths, and possibly even signs) which do not conform to the pyrochlore lattice connectivity, could significantly alter such a frustrated and delicate ground state. We therefore expect differences to exist in the magnetic specific heat, and ultimately the observed magnetic ground state.

%
%

\section{\label{sec:level1}Conclusions}

We have shown, using neutron powder diffraction, that a single crystal of Yb$_{2}$Ti$_{2}$O$_{7}$ prepared by the floating zone mehod using \emph{purely stoichiometric} starting material can be most accurately described by a 2.3\% stuffed pyrochlore model.  This model is an improvement over the purely stoichiometric model in three ways: it accounts for discrepancies in relative intensities of several Bragg peaks, it naturally explains the observation of an increased lattice spacing in the crystal compared to the sintered powder, and it produces a small decrease in R$_{wp}$, indicating a better fit in the Rietveld refinements.  Other models, such as titanium vacancies with no ytterbium substitution or an anti-site disorder model, do not meet all of these criteria.  

This study was motivated by the known variability of the specific heat anomaly near 200mK in single crystal samples of Yb$_{2}$Ti$_{2}$O$_{7}$.  Our results indicate that excess moments are present in at least one single crystal prepared by the standard method.  The excess moments appear to create an overall increase of anti-ferromagnetic interactions in the system, consistent with other studies of stuffed rare earth titanates.  The spatially random variation of stuffed moments is likely to give rise to spin-glass like behavior in this frustrated system, which could explain the observation of broadened specific heat features in place of a sharp specific heat anomaly that is observed in the sintered powder samples.  The room temperature lattice parameter could, in principle, be used to estimate the level of stuffing before performing more time-intensive experiments such as  low-temeprature specific heat measurements.  We hope this work will encourage more detailed characterizations of both single crystal and powder samples of Yb$_{2}$Ti$_{2}$O$_{7}$.  In particular, it would be of interest to investigate the role of \emph{intentional} (controlled) stuffing on the low temperature magnetic properties of the crystals.  

The authors would like to acknowledge support from P. Dube at McMaster University and K. Page at LANSCE.  K.A.R. and B.D.G. were partially supported by NSERC of Canada.    K.A.R. acknowledges T.M. McQueen and J.E.Greedan for very helpful discussions.

\end{document}